# Reducing Gaze Distraction for Real-time Vibration Monitoring Using Augmented Reality

Elijah Wyckoff[1], Marlan Ball[2], Fernando Moreu[2*]

[1] *Department of Mechanical Engineering, University of New Mexico, Albuquerque, NM 87106, United States*

[2] *Department of Civil, Construction and Environmental Engineering, University of New Mexico, Albuquerque, NM 87106, United States*

[*] *Corresponding author, fmoreu@unm.edu, 210 University of New Mexico, Albuquerque, NM 87131-0001.*

Operators want to maintain awareness of the structure being tested while observing sensor data. Normally the human's gaze shifts to a separate device or screen during the experiment for data information, missing the structure's physical response. The human-computer interaction provides valuable data and information but separates the human from the reality. The sensor data does not collect experiment safety, quality, and other contextual information of critical value to the operator. To solve this problem, this research provides humans with real-time information about vibrations using an Augmented Reality (AR) application. An application is developed to augment sensor data on top of the area of interest, which allows the user to perceive real-time changes that the data may not warn of. This paper presents the results of an experiment that show how AR can provide a channel for direct sensor feedback while increasing awareness of reality. In the experiment a researcher attempts to closely follow a moving sensor with their own sensor while observing the moving sensor's data with and without AR. The results of the reported experiment indicate that augmenting the information collected from sensors in real-time narrows the operator's focus to the structure of interest for more efficient and informed experimentation.

## 1. Introduction

Researchers quantify the response of structures by measuring and observing vibrations. Acquiring smart sensor data in real-time enables operators to predict failures and make informed decisions on maintenance [1]. This is enabled by IoT technology which is used for wireless sensor networks (WSN) for environmental sensing [2]. Researchers need to track vibration levels to prevent damage to sensitive machines, but current technology does not allow for a researcher to work freely without constantly checking a computer monitor [3]. Smart Infrastructure wireless sensors are useful for their reliability, low-cost, low-power and fast deployment characteristics [4]. Wireless sensor networks are used for monitoring and assessing vibration risk in historical buildings and cultural sites [5]. Forming a network of wireless sensors supports the gathering of data and decision making before, during, and after a crisis event. A wireless sensor network in Torre Aquila proved the system is an effective tool for assessing the tower stability while delivering data with loss ratios <0.01% with an estimated lifetime over one year [6]. Often data acquisition occurs prior to processing in wireless sensor systems for structural health monitoring (SHM), which is why researchers have explored implementing real-time wireless data acquisition on the Imote2 wireless sensor platform [7]. Researchers have also developed a vision-based tracking method to detect damage to a structural system using cameras already installed in the system [8]. Wireless and remote sensor systems are optimal for efficient and reliable data feedback, but there remain challenges for users to see real-time data. Open challenges remain that would be beneficial to explore in human-sensor interfaces.

AR is useful to researchers in informing of real-time data. AR has been used to augment different types of wireless sensor data through IoT technology [9]. Researchers augmented displacement data collected by smart sensors, however these values were first recorded and stored in a database before they were graphed in AR [10]. Researchers have also developed a human-machine interface which organizes metadata and provides actionable information by visualizing data about the built environment both on and off-site using AR [11]. Ballor et al. investigated using AR in infrastructure inspections where the framework uses the headset's sensors to capture a high-resolution 3D measurement of the infrastructure [12]. This can be used to analyze the state of the structure over time and track damage progression. AR has been used for SHM including detecting heat emitted from electronic equipment [13]. Wang et al. presents two Mixed Reality and AR systems and their application scenarios for the construction industry [14]. This study showed how these technologies can be integrated into heavy construction operations and equipment management, and they are emphasized for their potential to reduce cost, time, and levels of risk by augmenting applicable events with digital content. Implementing automated driving suffers from a problem with lack of trust and user acceptance, and AR technology exists as a solution to mitigate these issues. The prospect of increasing user acceptance and trust by communicating system decisions through AR is investigated by quantifying user acceptance using the Technology Acceptance Model [15]. AR for manufacturing training, specifically for welding, is evaluated using the Technology Acceptance Model to understand how welders perceive its practicality and ease of use [16]. AR has a wide range of uses making it a valuable tool for SHM, and this research seeks to develop a framework for the direct augmentation of live vibration data.

Gaze distraction is an important obstacle to consider in experimental work, and AR is used to address this issue. According to a review of AR technology, an estimated 80% to 90% of the information humans receive is through vision [17]. The ability to absorb and process information is limited by our mental capacity, and the same study examines how AR can reduce this cognitive load. Each mental task we undertake reduces the capacity for other, simultaneous tasks. AR technology is applied to vehicle operation using AR heads-up displays to lay navigational images directly over what the driver sees through the windshield [18]. This research proves how this can reduce the mental effort of applying the information, and it prevents gaze distraction because the driver focuses their attention on the road. AR is also applied to robot teleoperation to reduce gaze distraction, where augmenting live video feed from the robot limits the user's view to pertinent information for safer, more controlled operation [19]. Reducing gaze distraction in vibration monitoring looks to manifest safer operation and higher cognition in the same way.

This paper leverages AR technology to allow researchers to directly interact with the real-world through steady real-time communication with WSN providing quantitative information. AR technology is used to consolidate information in the user's view so that inspectors receive information regardless of where they are looking or positioned in the real-world. Traditional methods of vibration monitoring include a device with a screen that displays data. The new interface has been explored in the domain of structural design since it is now possible to interface the structural responses with holograms and other models permitting the researcher to quantify structural dynamics in the augmented interface. The interface includes a LEWIS5 (Low-cost Efficient Wireless Intelligent Sensor) which is an Arduino Metro M4 microcontroller equipped with an accelerometer to measure vibrations wirelessly. This data is sent over WiFi using TCP connection to the Microsoft HoloLens Gen 2 headset, where acceleration values are plotted real-time in the user's field of view. The proposed application is validated by a series of experiments testing a human's ability to react and maintain awareness of reality with and without AR. The human attempts to recreate the motion a moving sensor with their own sensor while also monitoring data, where the human's sensor data and eye movement data are collected. This work is innovative in human-structures interfaces, and it enables a new mode of sensing dynamics in real-time.

## 2. Framework

*2.1 Motivation*

Out of the five senses humans receive an estimated 80-90% of information from vision [17]. Understanding where information is best perceived by vision is important in this research. According to Younis et al. [20], central vision has the highest sharpness visually and is where humans pay the most attention to objects of interest. Human vision perceives a visual field of more than 200° diameter horizontally and 125° vertically, but this research is primarily interested in central vision which makes up an area of about 13° around the area of fixation [20, 21]. This field is modeled below in Figure 1. This research seeks to quantify the reduction in gaze distraction by tracking the area covered by the human's eyes with and without the aid of AR.

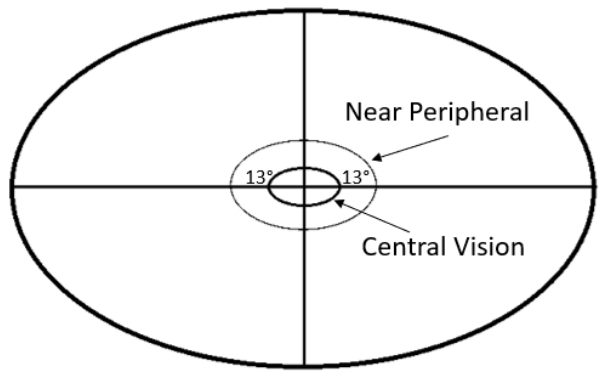

Figure 1: Model of central vision in human perception [20]

This project is developed based on a theory of human-structure interfaces. Researchers are interested in measuring vibrations and are informed by the device that receives the sensor feedback. If the device receiving sensor data is an AR headset, information can be relayed directly to the human [10]. This theory proposes that humans can be better informed and maintain better awareness of reality if they directly receive information on nearby structural response. Andersson et al. demonstrate AR in human-robot interaction, proposing improved training, programming, maintenance, and process monitoring by augmenting information [22]. Figure 2 illustrates vibration monitoring where it is necessary for the researcher to be present for experimentation. In this setup, the researcher monitors real-time vibration data collected from sensors secured to a frame. The researcher maintains focus on the suspended mass while a shaker generates excitations. Typically, data is recorded and plotted on a computer screen which requires the inspector to focus their attention on either the data or the structure. Monitoring both the data and the structure becomes difficult when the computer screen obstructs the researcher's view. The user also depends on the location of the computer for information, as it is inefficient and inconvenient to hold and carry around. This introduces potential issues with safety and control.

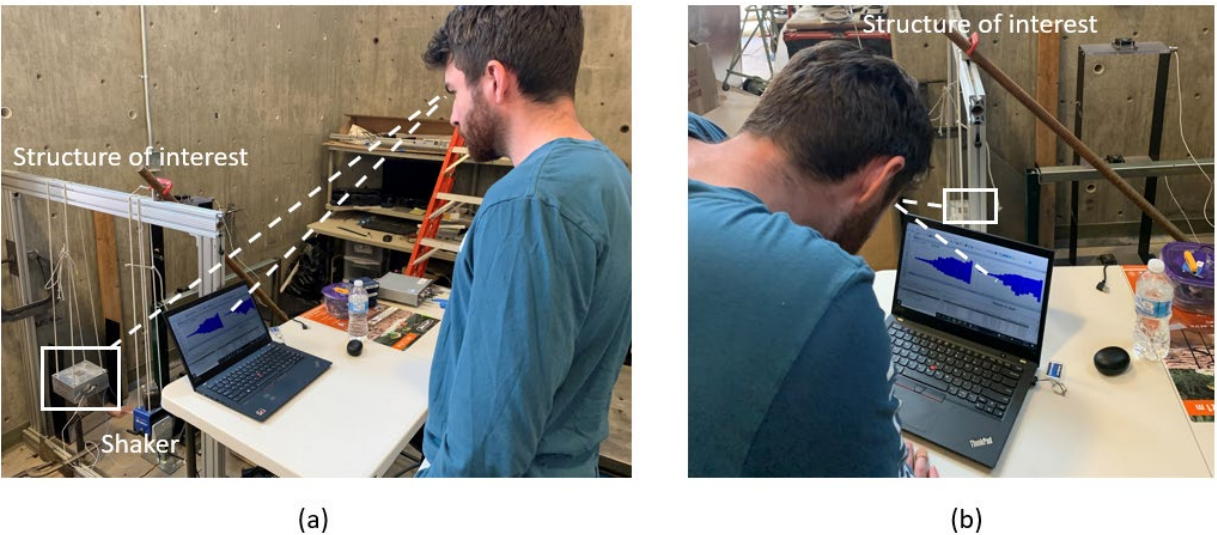

Figure 2: (a) Side view of the researcher's gaze while monitoring vibrations; (b) View from behind the researcher demonstrating obstruction by the screen displaying data.

*2.2 Proposed model*

By augmenting the plot of the live acceleration data, a loop between human and reality is formed that eliminates gaze distraction as a barrier to vibration monitoring. Figure 3 illustrates gaze distraction as a barrier and Figure 4 shows the proposed model aided by AR. The user receives direct information on reality via the augmented plot of live data in the AR headset thereby improving cognition of structural response while maintaining an area near central vision. In the framework of this research a user reacts to data by attempting to synchronize the acceleration of a handheld sensor with a moving sensor.

*2.3 New interface*

The new interface combines hardware and software to improve human cognition of sensor information. A connection between the sensor and user is formed by augmenting feedback in the user's vision, as shown in Figure 4. The AR headset is used to augment information in the form of holograms while maintaining awareness of the structure. In the proposed application acceleration data is plotted as a holographic chart in the user's view.

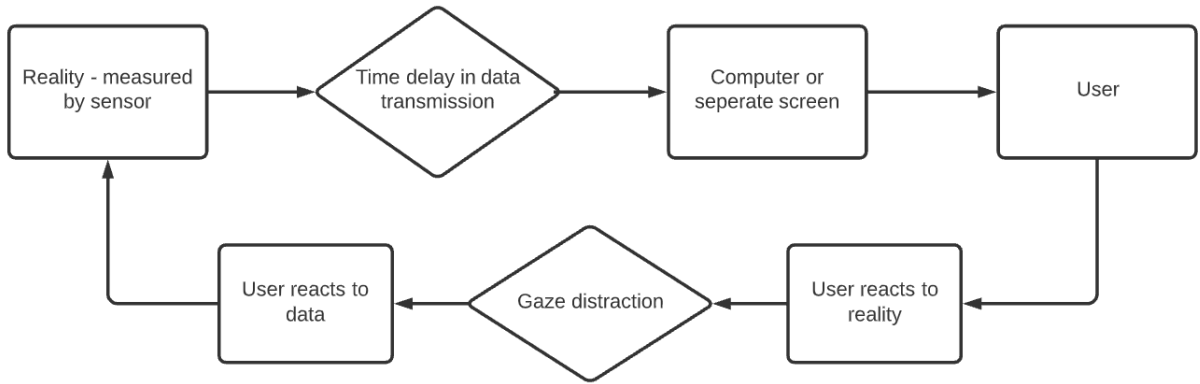

Figure 3: Current Model

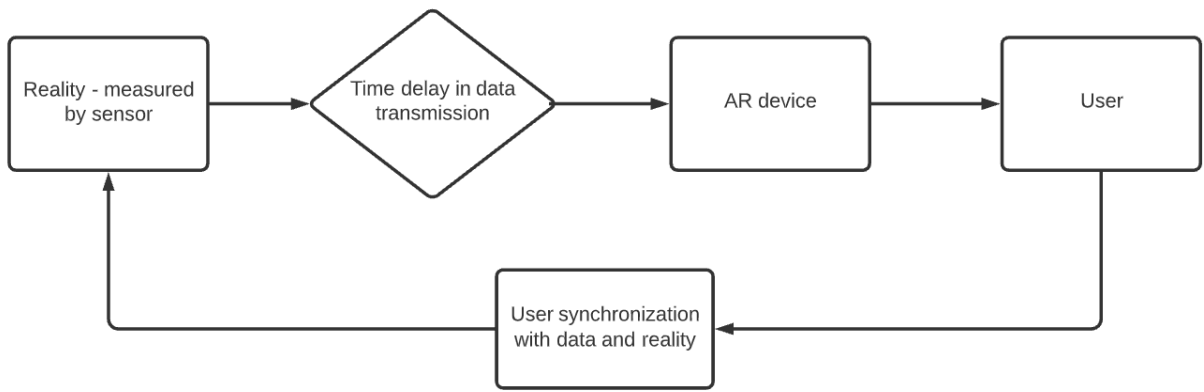

Figure 4: Proposed model

## 3. Hardware

AR blends interactive digital elements with a real-world environment. Holograms are generated by a computer and super-imposed onto the real-world environment, which allows the user to interact with the merged environment. This is enabled by a device that creates an AR environment via optical see-through display. The AR headset is a head mounted display that allows for contact free operation by hand gestures and voice commands.

*3.1 Augmented Reality Device Selection*

There were several factors to be considered in selecting an AR device for use in this research. These include the headset's sensing platform, system, display and interface, and general properties including weight, durability, battery life, price, and availability. Mascareñas et al. [23] gives an overview of these considerations used to make the device selection for this project. It was also important to consider the device manufacturer because development of AR applications varies depending on the platform. The system considerations include the processing unit, Random Access Memory (RAM), system on a chip (SoC) and the device's storage. Display capabilities include the resolution, field of view, aspect ratio, and refresh rate.

*3.1.2 Microsoft HoloLens 2 Selection*

Considering all the device selection criteria, the Microsoft HoloLens 2 headset was selected for development and application deployment in this project over the HoloLens first gen. The HoloLens 2 is the more expensive option but is the best AR device in terms of performance. Moreu et al. [24] summarizes the advantages of the selected device with a comprehensive breakdown of its features and capabilities. The Microsoft platform allows for Universal Windows Platform (UWP) development which is supported in Unity. A significant change in the HoloLens 2 from the first generation is the move from an x86 processor to an ARM-based platform for higher performance and power efficiency [25]. The field of view in HoloLens 2 is also improved, up to 52 degrees from 35 degrees in the first gen. Additionally, the HoloLens 2 enables eye tracking and hand tracking as opposed to the limited gesture tracking of the first gen HoloLens. A more detailed breakdown of the HoloLens 2 specs from Microsoft [26] is included in Table 1.

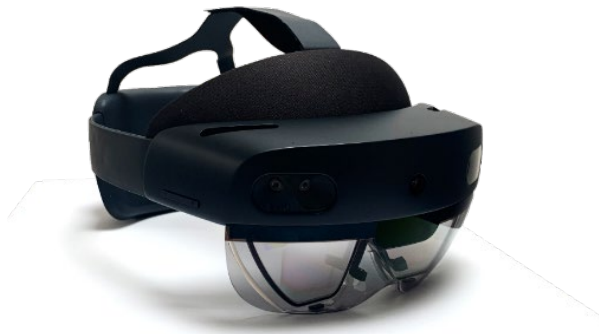

Figure 5: AR Headset – Microsoft HoloLens 2

*3.2 Sensing Platform*

This section describes the sensing platform developed for detecting and recording vibratory data. The sensing platform is developed to read acceleration data in a triaxial coordinate system as a wireless SHM system. This is done with a Low-cost Efficient Wireless Intelligent Sensor, abbreviated as LEWIS5. The LEWIS5 sensor is built by combining a WiFi shield and microcontroller with a triaxial accelerometer.

*3.3 LEWIS5 and its components*

This section provides an overview of the individual components needed to construct the sensor and includes a price breakdown to show the low-cost aspect of the sensor. A description and price point of each component is included in Table 2. The sensor connects via WiFi but requires a power source hooked up via micro-USB. The physical components are shown in Figure 5 and the fully assembled sensor is labeled in Figure 7.

Table 1: HoloLens 2 relevant features [26]

| Microsoft HoloLens 2 | |
|---|---|
| **General** | |
| Field of view | 52 degrees |
| SoC | Qualcomm Snapdragon 850 Compute Platform |
| Resolution | 2k 3:2 light engines |
| Storage | 64-GB UFS 2.1 |
| Weight | 566 g |
| Battery life | 2-3 hours active use |
| Connectivity | WiFi, USB Type-C, Bluetooth |
| Software | Windows Holographic Operating System, Microsoft Edge, Dynamics 365, 3D Viewer |
| **Sensors** | |
| Hand tracking | 4 visible light cameras |
| Eye tracking | 2 IR cameras |
| Depth | 1-MP time-of-flight depth sensor |
| IMU | Accelerometer, gyroscope, magnetometer |
| Camera | 8-MP stills, 1080p30 video |
| Microphone and speakers | 5 channels, spatial sound |

Table 2: Sensor breakdown

| Part | Description | Manufacturer | Price |
|---|---|---|---|
| Arduino Metro M4 Express | Microcontroller | Adafruit | $27.50 |
| Arduino Airlift WiFi Shield | Shield + WiFi co-processor | Adafruit | $14.95 |
| MMA8451 | Triaxial Accelerometer | Adafruit | $7.95 |
| Headers | Connectors | Sparkfun | $1.50 |
| Jump wires | Connectors | Sparkfun | $1.95 |
| **Total Cost** | | | **$53.85** |

*3.3.1 Metro M4 Express*

The Metro M4 Express is a 32-bit microcontroller with the ATSAMD51 microchip [27]. The Cortex M4 core runs at 120 MHz with floating point support. The board is powered via micro-USB or barrel jack connection. The board has 25 general purpose input/output pins, including 8 analog in, two analog out, and 22 PWM outputs. The pins can collect information from sensors for use in this project. It also includes a 2 MB Quad-SPI Flash storage chip which reads and writes programs from Arduino. The board is flexible, efficient, and affordable making it a good option for this project.

*3.3.2 Airlift WiFi Shield*

The Airlift WiFi Shield allows the use of the ESP32 chip as a WiFi co-processor [28]. The Metro M4 microcontroller does not have WiFi built in, so the addition of the shield permits WiFi network connection and data transfer from websites as well as the sending of socket-based commands. The shield includes a microSD card socket used to host or store data. The shield is connected to the microcontroller with stack headers. In summary, the WiFi Shield is necessary for wireless capabilities.

*3.2.3 MMA8451 Accelerometer*

The triple-axis accelerometer used for this project is the high-precision MMA8451 with a 14-bit Analog-to-digital converter [29]. The accelerometer is used detect motion, tilt and basic orientation designed for use in devices like phones and tablets. For the purpose of this project the accelerometer is used to detect motion, especially vibrations. Its usage range varies from ±2G up to ±8G which ideal for its application to this project.

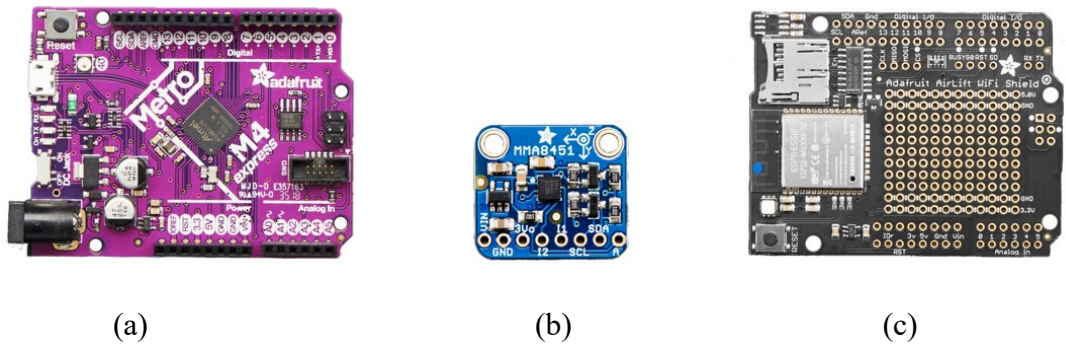

(a)  (b)  (c)

Figure 6: Components of the LEWIS5 sensor. (a) Metro M4 Express; (b) MMA8451 Accelerometer; (c) Airlift WiFi Shield

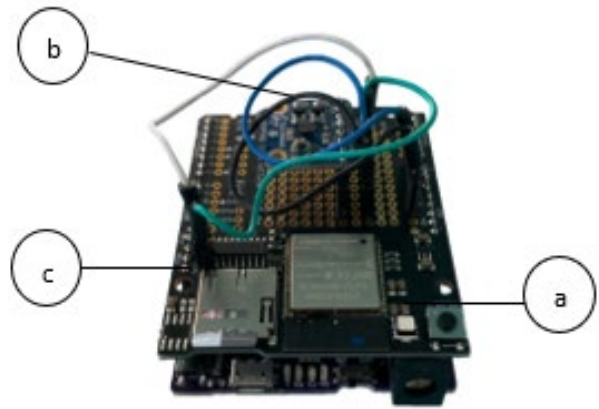

Figure 7: LEWIS5 sensor full assembly

**5. Software and Development**

Programming and development of the AR application is done in Unity version 2018.4.19f1 taking advantage of the Mixed Reality Toolkit (MRTK) from Microsoft. The MRTK is applied to a scene built in the Unity application to configure the scene for AR use. The application is developed for Universal Windows Platform which allows deployment to the HoloLens 2. The programming platform is Visual Studio 2019, and the Unity scene programming is written in C#.

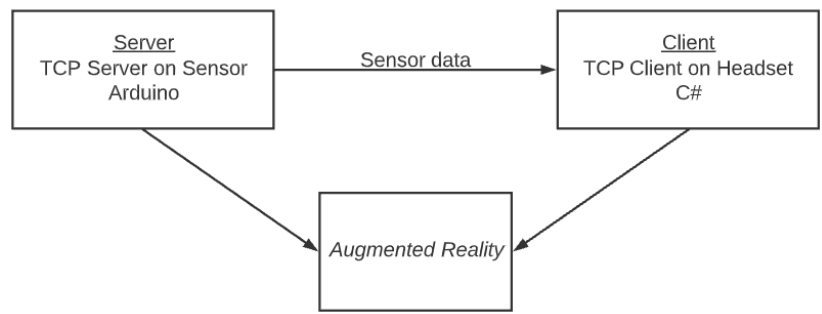

Figure 8: Software components

## 5.1 Arduino Programming

The sensor programming was performed in the Arduino IDE, an open-source software environment that is written in Java and based on Processing and other software. This program facilitates the writing and uploading of code for any Arduino board as well as other compatible systems.

### 5.1.1 Server Creation

The WiFiNINA library is available for download in the Arduino IDE. This library enables the LEWIS5 sensor to be set up as a Transmission Control Protocol (TCP) server in the Arduino code. The board connects to a nearby WiFi network and accepts incoming connections on the port it is listening on. If the network is private the Arduino code includes a secret tab with the network name and password. Existing scripts for the MMA8451 accelerometer were modified to read, print, and send the acceleration data at a sampling rate of 20 points per second. The Arduino Serial Monitor prints the SSID of the network it is connected to and confirms the WiFi connection. The board will then wait for a client connection before it begins printing the accelerometer values. The Serial Monitor window begins auto-scrolling with the three columns of acceleration data once a client successfully connects. There is a slight time delay in the augmented plot of sensor data induced by the network connection, which was investigated by the researchers in a series of 12 tests. The tests were conducted on a mobile hotspot which is used as the WiFi network for the experiment section of this paper. It was discovered that the average time delay was about 0.26 seconds on the hotspot, which is taken into consideration when reviewing results.

## 5.2 Unity Development

Unity Game Engine version 2018.4.19f1 was used for cross-platform development as it supports open-source programming for headsets and mobile devices. The Unity scene is configured with Microsoft's MRTK library to support the AR features of the application. The toolkit includes default scripts for necessary features in the HoloLens such as gestures, commands, and interface features.

*5.2.1 Client Connection*

Modified code from Timur Kuzhagaliyev [30] is implemented for connecting the HoloLens and Unity to sockets. The process implements a TCP client that works for development in the Unity editor as well as for development in UWP on HoloLens. Functions in the Windows Sockets namespace System.Net.Sockets are used to connect the HoloLens as a client to the open port on the sensor's server.

*5.2.2 Graph development in Unity*

The graph of the live data is developed as a scatter plot, which was chosen as the most effective and efficient solution. The graph is developed based on a tutorial from Catlike Coding [31]. Points at each appropriate coordinate are generated by Unity's default cube game object, which are color coordinated based on x, y, and z acceleration. Each data point is graphed as a small 3D cube for visual feedback. The Transform component is used to position each individual cube, which are variably instantiated as clones. Vector3 creates a 3D vector which defines the position of each cube. The incoming data is parsed to define each point of Vector3. At any given time there are 100 cubes generating the data lines in the display. This is defined by the resolution set in Unity, as the number of cubes is set to the value of the resolution. These cubes are connected with a LineRenderer command that makes the displayed data appear as a line chart rather than individual cubes. The graph updates with each frame meaning the cubes are adjusted as time progresses, defined by the function $f(x,t)$.

*5.3 Application Development*

The problem addressed in the following section is the lack of a user-friendly interface for an AR application for live accelerometer feedback. The previous model was a bare plot of the three acceleration lines. The developed interface provides the necessary inputs for commands including client connection and disconnection and graph initiation and shut down. The interface also includes a means of providing the user with a warning system for the breaching of a user-specified threshold value. Figure 9 illustrates the details of application development in the form of a flowchart.

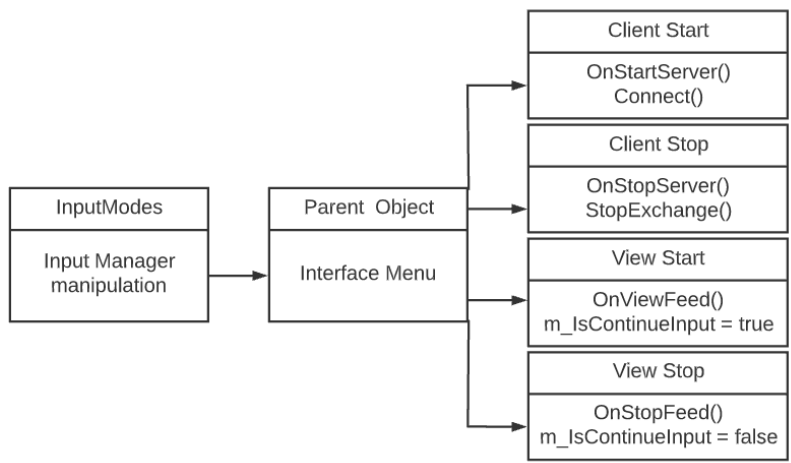

Figure 9: Application flowchart

## 5.4 Interface menu and functions

This section presents the interface of the AR application and explains the function of its unique features. The full view of the interface is shown in Figure 10. The application interface consists of six different buttons with specific functionality. The following subsections contain a detailed explanation of these functions and their use.

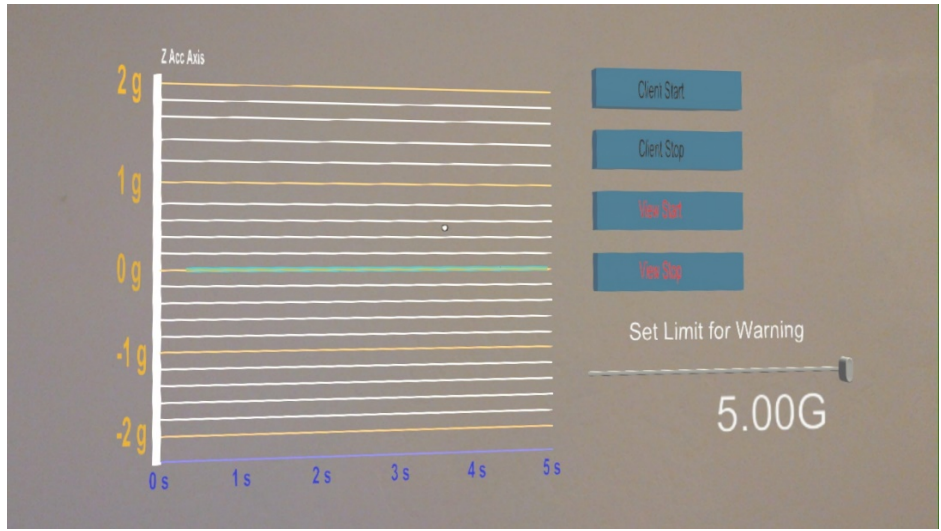

Figure 10: Interface menu and graph

### 5.4.1 Client Start

Client Start connects the client to the server via TCP. In the context of the application, the computer running the Arduino program acts as the server and the device running the AR application is the client. The Unity code requires the IP Address of the Arduino board, and the Unity code and Arduino code are set up on the same port.

### 5.4.2 Client Stop

Client Stop closes the client connection to the server. The live data feed flattens to zero and the Arduino program must be rerun to initiate another connection.

### 5.4.3 View Start

This button initiates the function ContinueInput. Incoming data from the server is parsed into x y and z vectors. This corresponds to the axes of the accelerometer. The graph plots the data from left to right as three color-coordinated lines. Data is converted to terms of the gravitational constant G. The x and y data are also offset so that the x line does overlap and hide the y line. Therefore, the graph axis is labeled as "Z Acc" for the purpose of the experiment as well as simplicity. Future work on the application will include the addition of x y and z axes selection.

### 5.4.4 View Stop

Stopping the view zeros out the three data lines but does not disconnect the client. The view may be resumed by selecting View Start again.

*5.5 Positioning the graph*

In the early development stage of the application the acceleration lines plotted at an arbitrary point in space. To verify accurate positioning of the horizontal axis lines the graph was developed using known input from an electrodynamic exciter. The exciter vibrates at a user-defined frequency to enable exact placement of the axis lines. The x axis represents values of time in seconds that are spaced according to the sampling rate. By measuring one second intervals the x axis labels were placed accordingly.

*5.5.1 Electrodynamic Exciter*

The SmartShaker Model K2004E01 electrodynamic exciter from The Modal Shop is a small, portable permanent magnet shaker with a power amplifier integrated in its base [32]. The excitation signal from a function generator is plugged directly into the BNC connector at the base of the shaker. The framework also includes a separate AR application which can be used to change the input to the shaker wirelessly. The SmartShaker provides up to 7 pounds pk sine force and is supplied with a DC power supply. Benefits of the shaker include the integrated power amplifier, easy mounting and positioning, and 10-32 threaded mounting inserts for payloads up to 2 lbs. The LEWIS5 sensor is mounted to the shaker by a 10-32 nylon stinger as shown in Figure 11.

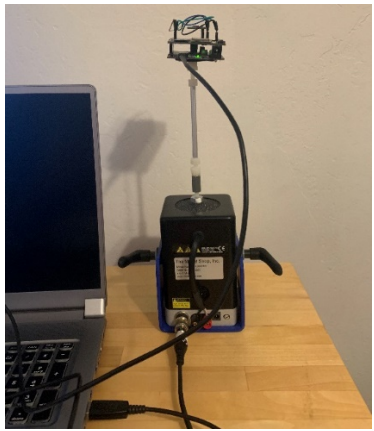

Figure 11: Graph development and verification – Sensor-Shaker Configuration

## 6. Investigating Reduced Gaze Distraction

*6.1 Experimental Objective*

To fully understand reality, humans receive information from the physical space while relying on sensors for data and information they cannot detect with their own senses. Researchers have examined human ability to tap their fingers at frequencies of 1, 2, and 3 Hz to investigate manual dexterity of elderly subjects [33]. For this research, a researcher is tasked with following a moving sensor with a second, handheld sensor while also maintaining awareness of the data received from the moving sensor. The moving sensor is run at 1, 1.5, 2, 2.5, and 3 Hz. The objective of the experiment is to measure the level of gaze distraction while monitoring and attempting to recreate vibration data with and without AR, where it is hypothesized that human has a better sense of reality when the data is augmented in their central vision. Quantifying the area covered by the user's eyes and the user's ability to follow a moving sensor provides a means of understanding the value of AR as a tool for data visualization and control. Figure 12 demonstrates the value of AR in reducing gaze distraction by modeling the primary area of interest and its proximity to central vision in the three experimental cases – reality, monitoring data with a device, and monitoring data with AR.

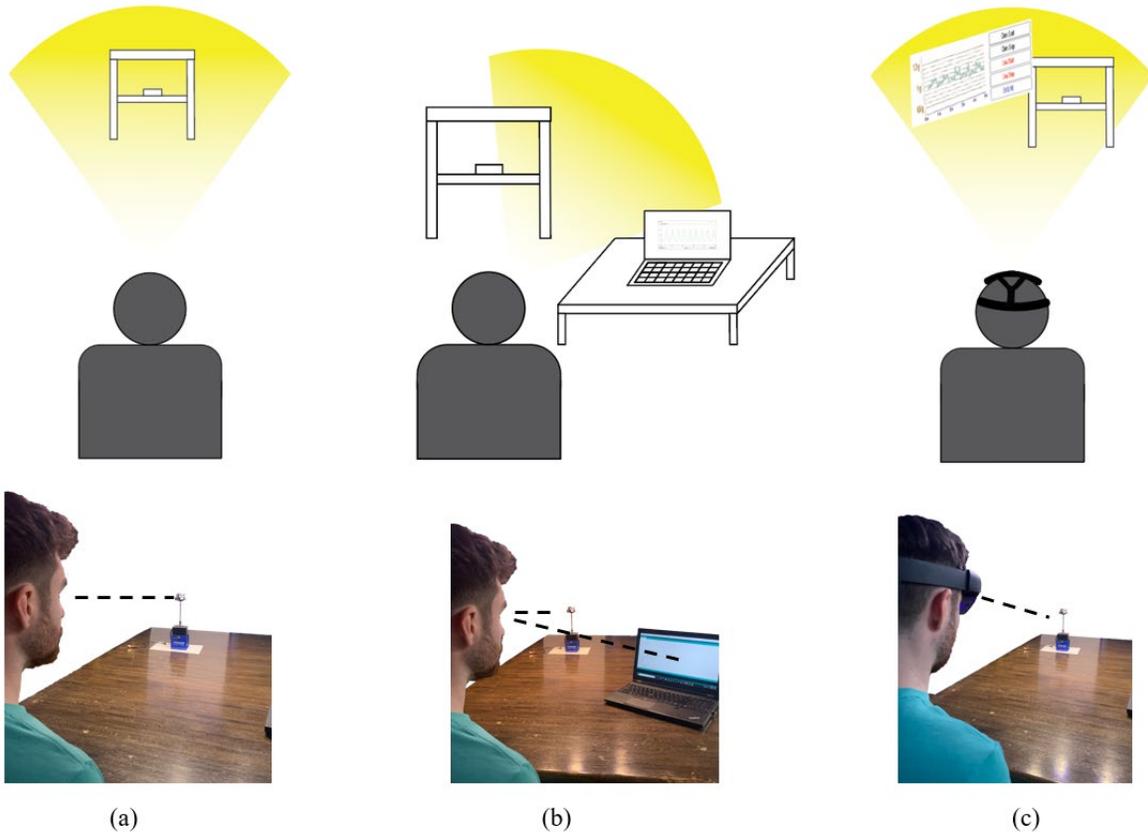

(a) (b) (c)

Figure 12: (a) The researcher maintains maximum awareness of reality in their central vision; (b) The area of interest is not fully in central vision when checking data; (c) The area of interest and data feedback are constrained to the HoloLens user's central vision.

*6.2.1 Experimental Setup and Procedure*

The experiment was set up with two laptop computers, two LEWIS5 sensors, a smart shaker, and the Microsoft HoloLens 2. One laptop computer provided power to the shaker sensor and the other laptop computer supplied power to the handheld sensor. The shaker sensor, the first laptop, and the HoloLens are connected to the mobile WiFi hotspot mentioned in Section 5.1.1 to send data from sensor to HoloLens and from HoloLens to MySQL database. The second laptop was also used to plot sensor data when measuring gaze distraction without AR. The researcher acting as the subject was positioned standing one meter from the sensor-shaker setup. The shaker was run at 1, 1.5, 2, 2.5, and 3 Hz where a second researcher and the subject synchronize the sensors with a vertical excitation. The researcher acting as the subject begins following the shaker sensor at their discretion for a period of approximately 12 seconds. They were also instructed to maintain awareness of the data while following the moving sensor. This generates a sinusoidal plot which can be compared to the plot of the shaker sensor data to obtain time delay. Additionally, the data can be analyzed in the frequency domain to determine how well the user was able to synchronize with the shaker sensor. This data is collected using the HoloLens 2 eye tracking API, which from a target of one meter can be plotted in terms of x and y coordinates with an accuracy of 1.56 cm [34]. The user must click a button in the application UI to begin eye tracking, thus the points at the beginning and end are removed during analysis. All analysis and plot generation are done in MATLAB. Figure 13 shows the experimental setup with plotted eye tracking and the MATLAB results of the human's eye movement.

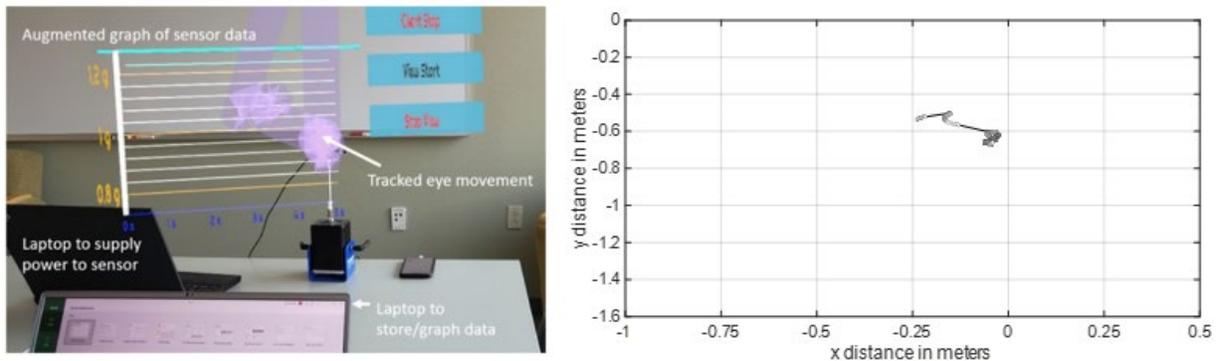

Figure 13: Experimental setup and example of eye tracking with AR graph and shaker-sensor

*6.2.3 Experimental Results and Analysis*

The eye tracking data is sent from the HoloLens to a MySQL database, which is then exported as a JSON file and converted to a string and parsed in MATLAB so that the data can be plotted. The start and end points are removed by reducing the range of the data. Each point has a three-dimensional coordinate, but this research is concerned only with the vertical and horizontal position of the eye movement. The string of data can then be graphed in MATLAB where each point is plotted and connected with a solid line representing the path of eye movement. The eye tracking data is sent along with time stamps which allowed the researchers to calculate an average sampling rate. The researchers are aware of the variable frame rate in AR applications and especially in AR applications communicating with devices like sensors. This influences the sampling rate of eye tracking data, and this is taken into account through a method of collecting the real sampling rate. For example, in five experiments the researchers determined the sampling rate by collecting eye tracking data while running the sensor plot in the same application for multiple iterations. The approximated sampling rate for the five experiments was 34. Eye tracking points for three experiments at 1.5 Hz are collected to demonstrate the importance of gaze distraction. Researchers conducted the same experiment at the three scenarios and collected the eye tracking points for approximately 50 oscillations. The time varied between 30 and 40 seconds depending on the experiment.

Figure 14 shows the results from the eye tracking while the human is trying to match the data by observing the experiment without any dataset. The results show that the area of eye tracking is very concentrated apart from four diagonals that can be attributed to the human's eyes drifting to the table. Nevertheless, the eye tracking data of this figure shows how the gaze distraction is minimized for the entire duration of the experiment.

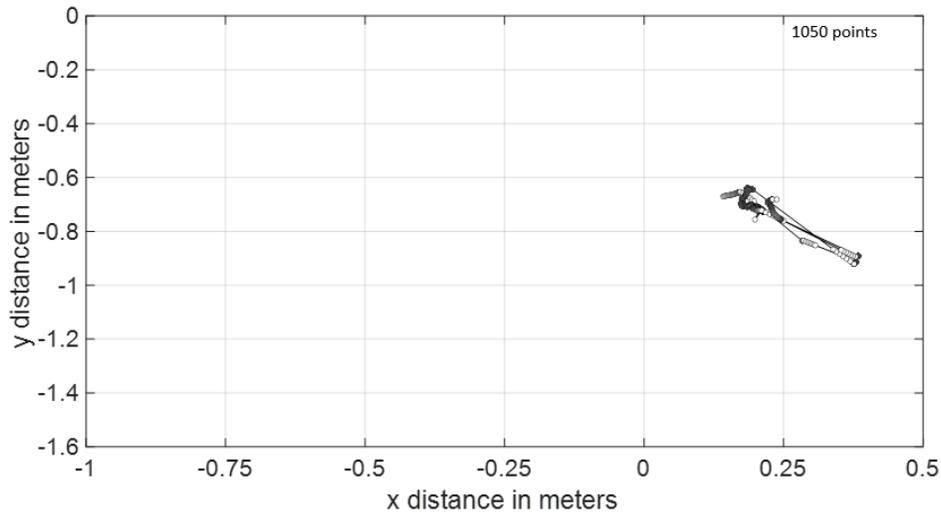

Figure 14: Eye tracking results while strictly monitoring the sensor

Conversely, Figure 15 shows the results from the eye tracking while the human is trying to match the moving sensor by observing the experiment while data is plotted on a laptop screen. The figure shows that eye tracking covers the space in between the screen and the moving sensor as the human attempts to maintain awareness of both. This depends on the positioning of the monitor, so results vary depending on the experimental setup. For the purpose of the experiment the laptop was in front of the human and 1 m from the shaker setup.

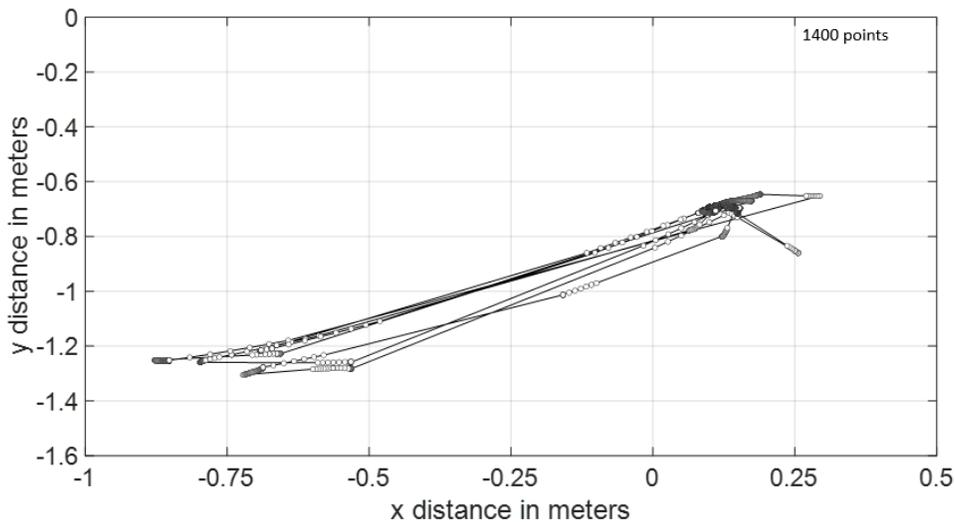

Figure 15: Eye tracking results monitoring data plotted on a separate screen

Figure 16 shows the results from the eye tracking while the human is trying to match the moving sensor while monitoring data in AR. The results show that the area of eye tracking is extremely concentrated with only one diagonal observed where the human's eyes drifted to the left side of the augmented plot. The eye tracking data is heavily concentrated because the hologram of the plotted data is augmented directly on top of the moving sensor.

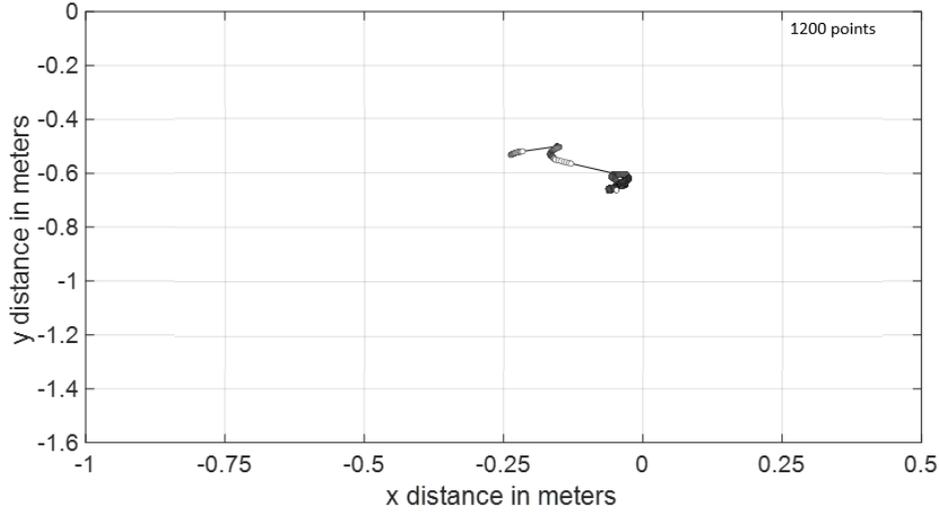

Figure 16: Eye tracking results with the AR plot

As expected, the eye tracking results shown in Figures 14-16 prove the inspector covers an area much closer to central vision than when monitoring data on a separate screen. These results help quantify the reduction in gaze distraction when monitoring an augmented graph of sensor data rather than a separate screen. The eyes drift 0.24 m from the primary area of focus (the shaker sensor) as opposed to covering 0.97 m of space outside of central vision when checking a separate screen. The human's eyes also drifted even when instructed to remain solely focused on the sensor, whereas the user did not get distracted with AR. The value lies in the results obtained with AR as the graph can be augmented on top of the area of interest without needing to be supported in some way or blocking the user's view, hence the minimal amount of eye movement observed in the results obtained with AR.

The sinusoidal plots of the handheld sensor and the shaker sensor are plotted from the recorded data according to the sampling rate of the sensor. The time vector for the plot is generated from known values of the length of the recorded data and the sampling rate. The peak-to-peak distance between each of the first 10 shaker and human excitations is recorded manually and the average is reported as the time offset for each test as per Equation 1. The shaker plot has slight dips that indicate the point at which the shaker briefly pauses at the top and bottom of its motion, and the peaks of the human's sensor movement are clearly defined. These are the points taken as $t_{shaker}$ and $t_{subject}$.

$$Time\ Delay = \frac{1}{N} \sum_{i=1}^{10} (t_{shaker} - t_{subject}) \qquad (1)$$

Figure 17 shows the time history of the first 10 excitations for each experiment, where the x axis is the time duration of the 10 excitations in seconds. The plots are normalized to include the first 10 excitations for each experiment, hence the x axis labels are removed and labeled as nondimensional time. Notably, the human's response was inconsistent in both synchronization and amplitude when monitoring the data on the laptop screen. The results at 1 Hz are the clearest example of the difference between monitoring the laptop screen and monitoring data in AR. The response aided by AR closely matches the shaker, whereas the response aided by the laptop screen is significantly off for the last nine excitations. The results aided by AR also display consistent amplitude for each of the individual experiments when compared to the "with screen" results, and the standard deviation of the amplitude of the peaks of 10 excitations is taken to examine this result.

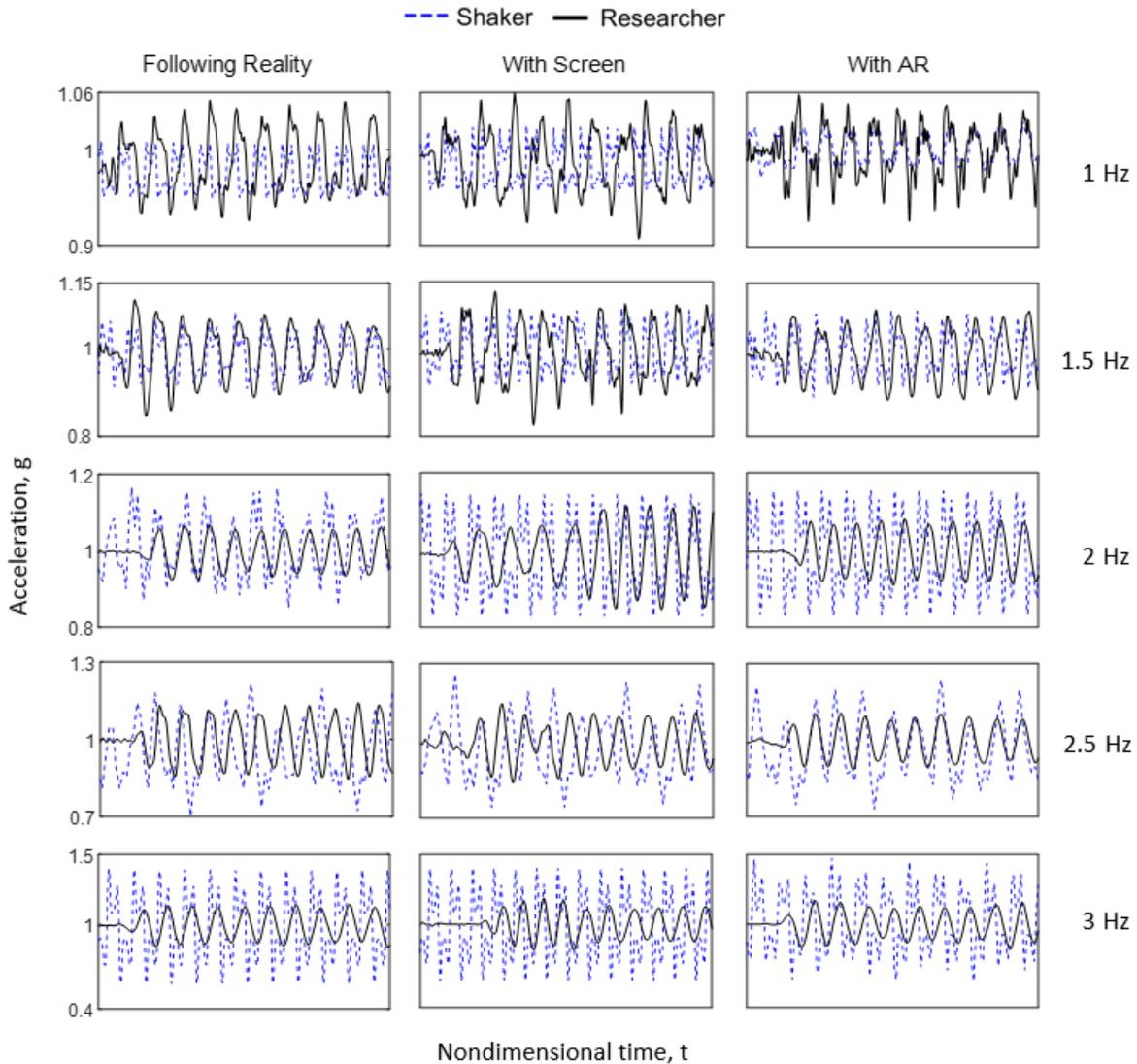

Figure 17: Time history of each experiment

Figure 18 shows each individual PSD generated for the signal in relation to the frequency of the shaker, which is indicated by the vertical black line. These results are used to understand how well the human synchronized with the moving sensor. Auto-spectral density estimates were generated for each single-input signal using Welch's method. This returns estimates at specified frequencies defined in the range of the sampling rate [35]. The truncation window is set to reduce uncertainties, where an integer multiple of 16 times the sampling rate is used to set the truncation window for each calculation [36]. Spikes in the PSD indicate that the signal is correlated with itself at regular periods, and thus indicate the spectra with the greatest effect [37]. This is done to determine the frequency of each signal, including that of the shaker since the shaker frequency cannot be assumed to be exact. The results for following the shaker while monitoring data on a computer screen, termed "with screen," indicate an asynchronous result in each PSD. Conversely, the PSD results with AR show that the human was able to generate a signal with a frequency close to that of the shaker sensor. The exact value of each offset is reported in Figure 20.

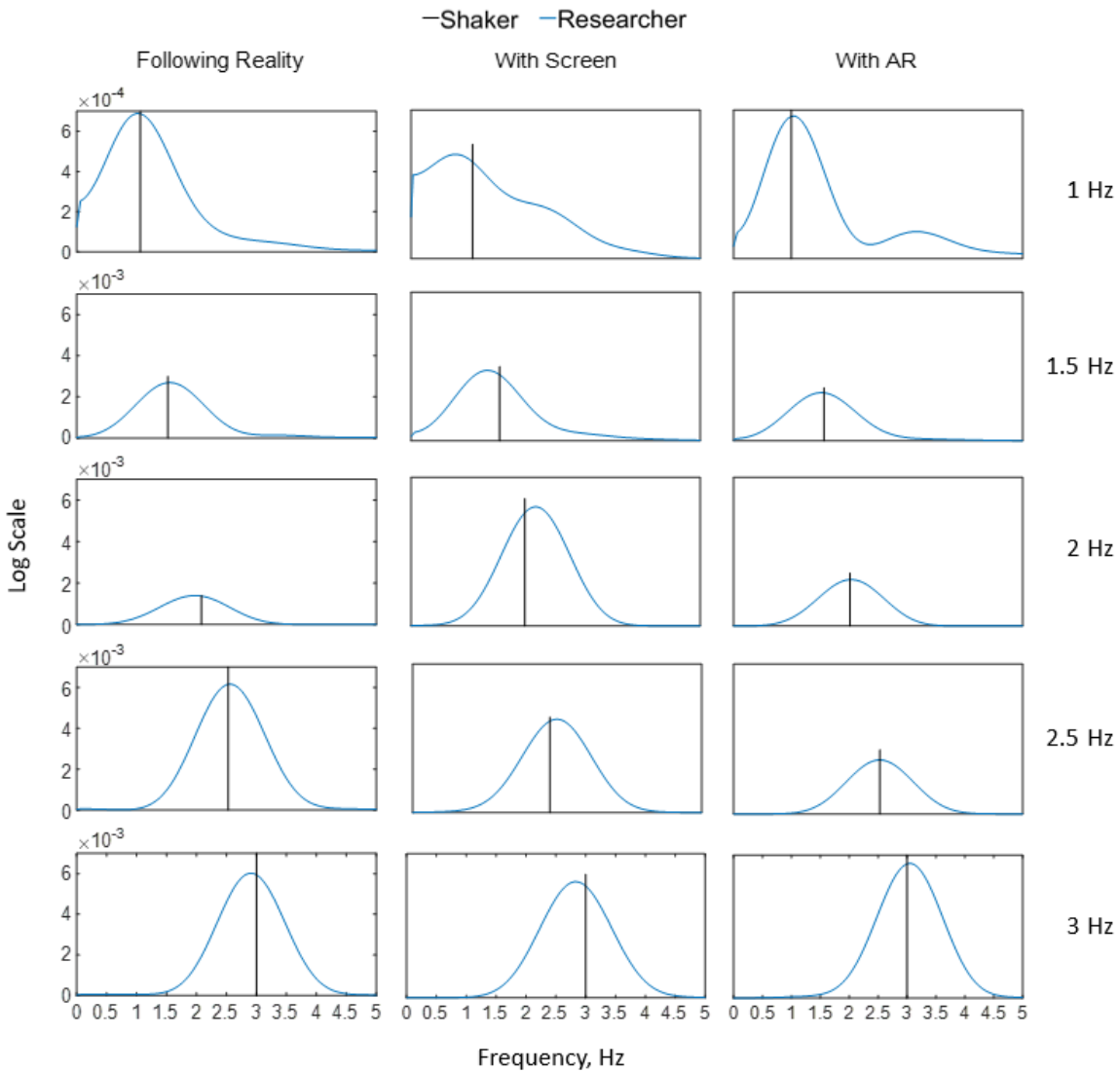

Figure 18: PSD of each experiment

Figures 19-21 display bar graphs of the reported results. The results are calculated from the range in which the human attempted to follow the shaker, with the first 10 excitations considered as the range for time delay calculations. Combining the eye tracking results with the results from the handheld sensor prove increased awareness of reality while using AR. Experiments at higher frequency were considered, however the human has difficulty recreating a faster response and the results are less valuable with shorter excitations. As expected, the human performed the worst when attempting to maintain awareness of data plotted on the computer screen. Figure 19 reports the average time offset between the response generated by the human and the response from the shaker sensor. The human struggled the most at 1 and 2 Hz with the separate screen, with an average delay of 0.31 and 0.3 seconds respectively.

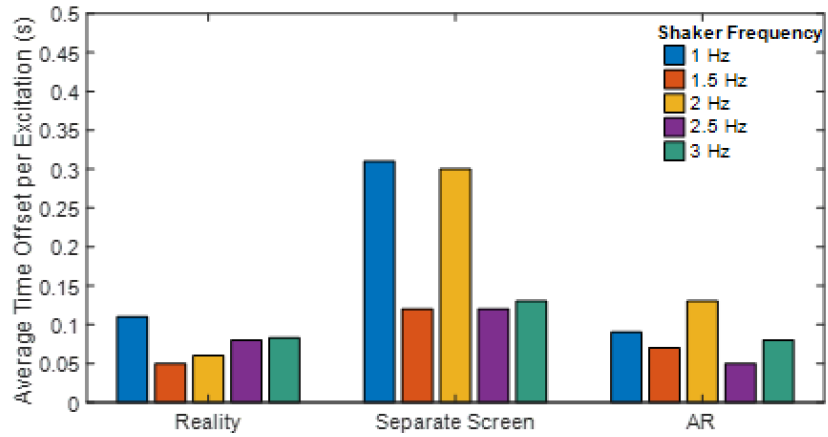

Figure 19: Results of time offset in user's attempt to follow moving sensor

Figure 20 shows the results of the human's synchronization with the moving sensor calculated from the PSD results of Figure 18. The human created a response with significantly worse synchronization and consistency when monitoring the computer screen. Conversely, they generated a frequency with less than a 0.1 Hz offset for each of the experiments with AR and reality. Notably, the human performed better with AR at 2 Hz than solely following reality and had very similar results at the other four frequencies.

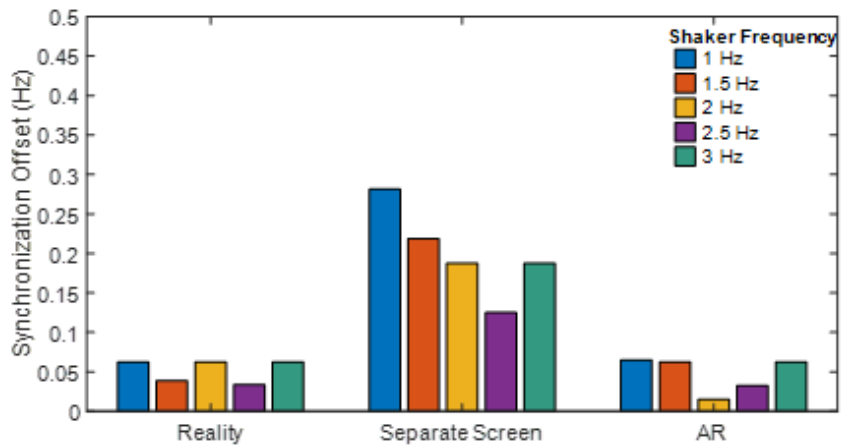

Figure 20: Results of user synchronization with moving sensor

Figure 21 displays the results for the standard deviation of the 10 peaks of the signal generated by the human. The human generated consistent amplitude at 1.5 Hz compared to the other two cases, however the standard deviation of the excitation peaks for the other four experiments was much higher in comparison. The human was more consistent with AR for each experiment with similar standard deviation compared to the results with reality.

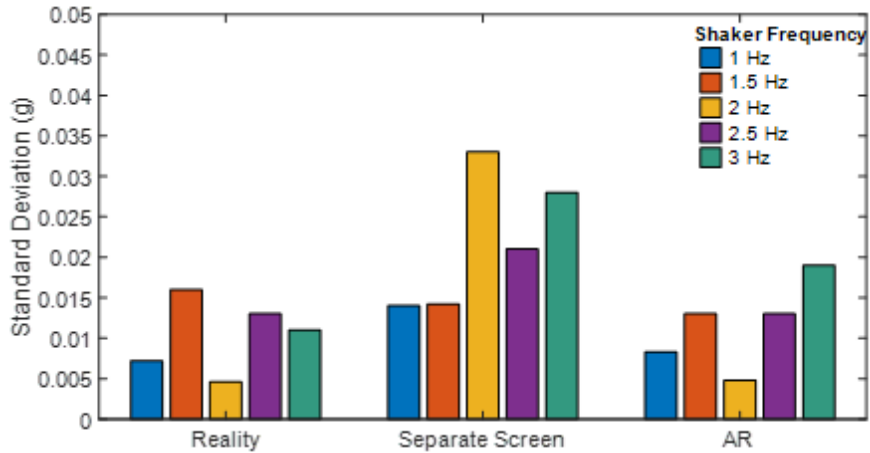

Figure 21: Results of user consistency in amplitude for the 10 excitations

From the combined results for time offset, synchronization and consistency it can be concluded that AR is an improved solution in vibration monitoring. Compared to the results of the case following reality, the results with AR are consistently in a similar range. This conclusion was expected as AR provides the ability to focus on both reality and data, whereas monitoring data with a separate device does not.

**7. Conclusions**

This paper developed and tested an AR application for live sensor feedback to reduce gaze distraction in vibration monitoring. An experiment was conducted to determine if augmenting data gives a human better awareness of reality by allowing the human to remain focused on the physical space. By tracking the human's eyes, an experiment proved that gaze remains close to the primary area of focus when monitoring vibration data in AR. Additionally, the human was able to use a handheld sensor to closely replicate the response of a sensor in the primary area of focus while maintaining awareness of the vibration data. Compared to the same test with the data shown on a separate screen, the human performed significantly better which demonstrates the improved sense of reality. This project has the potential to expand upon the current model for the inclusion of multiple sensors, different types of sensing devices and states, and other information pertinent to an inspector's interests. This implementation of AR technology reduces gaze distraction in vibration monitoring and allows inspectors to monitor both the physical space and the collected data for awareness and safety.

**Acknowledgements**

The financial support of this research is provided in part by the Air Force Research Laboratory (AFRL, Grant number FA9453-18-2-0022), and the New Mexico Consortium (NMSGC SUB AWARD NO. Q02151). The authors would like to extend thanks to Dr. Chris Petersen and Dr. Derek Doyle for their support and feedback in the project.